\documentclass[eqsecnum,,showpacs,aps]{revtex4}  
\usepackage{amsmath}
\usepackage{color}
\usepackage{graphicx}
\newcommand{\be}{\begin{equation}} 
\newcommand{\ee}{\end{equation}} 
\newcommand{\bea}{\begin{eqnarray}} 
\newcommand{\eea}{\end{eqnarray}} 
\newcommand{\mb}{\mathbf}

\newcommand{\mA}{\mathcal{A}}
\newcommand{\mT}{\mathcal{T}}

\newcommand{\del}{\delta}

\newcommand{\figph}
{\begin{figure}[htbp]
        \centering
        \includegraphics[angle=270,width=8cm]{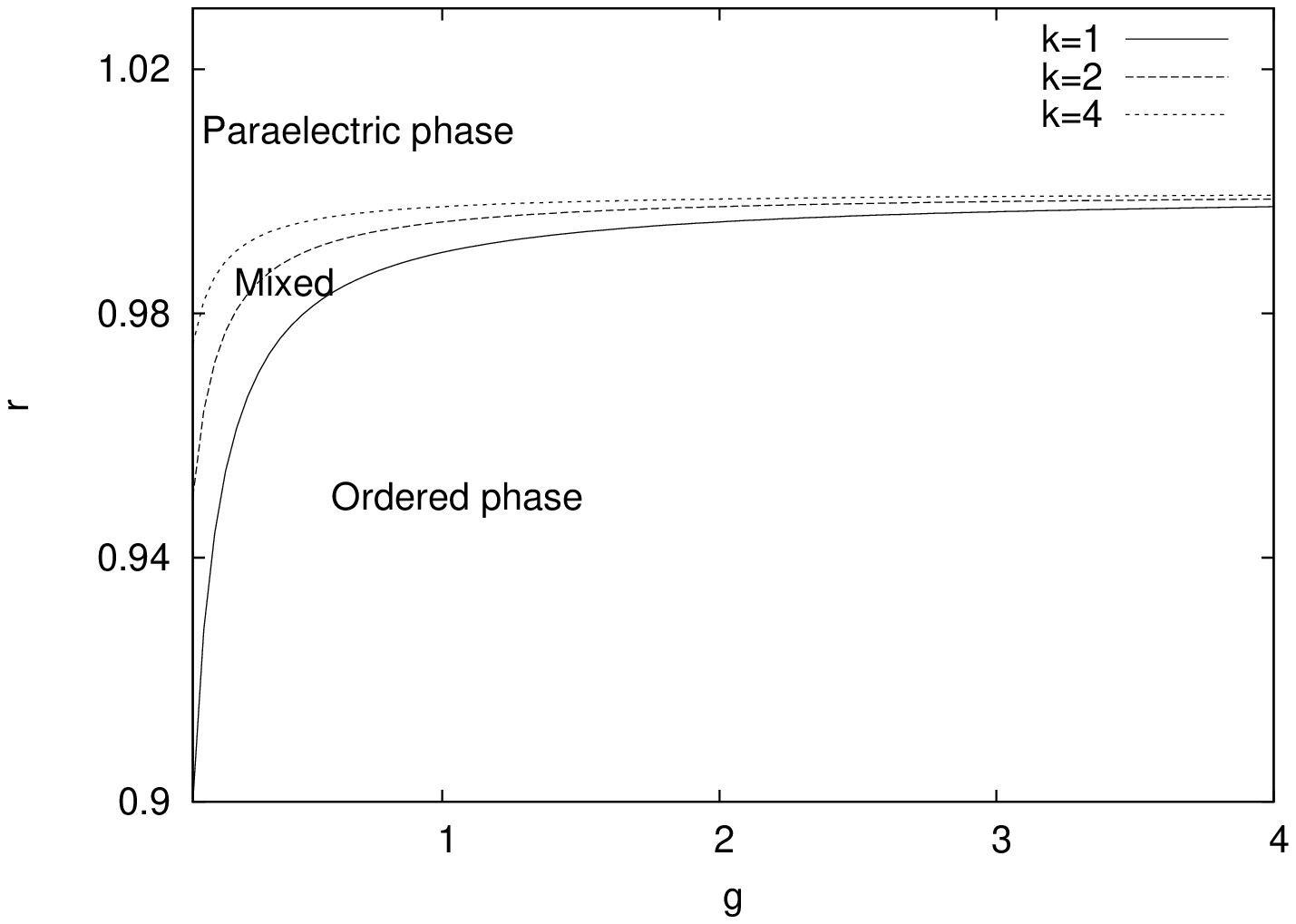}
\caption{Schematic phase diagram at zero temperature in $g-r$ plane for $k=1, 2$ and $4$ using eqn. \ref{rc}. Above a certain value of $g$ there is an ordered phase for $r>0$ . Moreover the region 
between  lines corresponding to various values of $k$ represents a mixed phase.}
\label{phase}
	\end{figure}
}

\begin{document} 
\title { On the possibility of mixed phases in disordered quantum paraelectrics}
\author{Nabyendu Das}

\email{nabyendudas@gmail.com}

\affiliation{Institute of Physics, Bhubaneswar 751005, India.}

\begin{abstract}
We present a theory of phase transition in  quantum critical paraelectrics 
in presence of quenched random-T$_c$ disorder using replica trick. 
The effects of disorder induced locally ordered regions and their slow dynamics are 
included by breaking the replica symmetry at vector level. The occurrence 
of a mixed phase at any finite value of disorder strength is argued. A 
broad power law distribution of quantum critical points and and its finite temperature consequences are predicted.  
Results are interesting in the context of a certain class of disordered 
materials near quantum phase transition.
\end{abstract}
%\keywords{quenched disorder, quantum phase transition, quantum paraelectrics}
\maketitle  
\section{Introduction} 
Following the recent progress in understanding quantum phase 
transitions in ferroelectrics\cite{I}\cite{Palova}\cite{II}\cite{Millis-1}\cite{Saxena} and the strongly correlated 
systems with disorder,\cite{Huse}\cite{Millis}\cite{Jones}\cite{Vojta} we make an 
attempt to understand certain features of quantum phase transition in disordered 
quantum paraelectrics. In general, a dilute amount of quenched impurity and 
associated disorder can create locally ordered regions even above the transition 
point of the corresponding pure system. Near a phase transition large size droplets become more probable. Moreover their slow dynamics becomes an important 
factor to determine the nature of quantum phase transition in a disordered system.
We address such issues in case of quantum phase transition in some dielectric materials, like  SrTiO$_3$, KTaO$_3$ etc. These materials are structurally similar to BaTiO$_3$, an well known text book example of ferroelectric material undergoing a ferroelectric transition at around $110$K. Such a transition accompanies spontaneous ferroelectric order below the transition temperature, divergence in static dielectric constant and softening of a transverse zone center optic mode right at  the transition temperature\cite{Shirane}. In spite of structural similarity with BaTiO$_3$, SrTiO$_3$ and KTaO$_3$ do not show any 
ferroelectric transition even at zero temperature in their pure form, rather end up with a very high, temperature independent, static dielectric 
susceptibility at low temperatures\cite{Muller}. The absence of a phase transition and the saturation of dielectric susceptibility at very high value in these materials also correspond to an incipient soft zone center transverse optic mode.  In recent times  a scenario is put forward that these materials  are near a ferroelectric quantum critical point and polarization fluctuations near such a quantum critical point forbids ferroelectric order and determines these unusual dielectric behavior\cite{I}. 
There is also very recent experimental works that supports this scenario\cite{Saxena}.
Extensive studies in this line took place during past several years. The role of polarization fluctuations in the low temperature regime in the suppression of ferroelectric transitions and the effects of dipolar interactions\cite{Millis-1}, pressure\cite{Sakudo} and impurity\cite{Muller-imp} were emphasized in earlier literature.  A relevant question in this line would be the effects of impurity and disorder in these systems. When an impurity  is introduced in such a system, it couples to local polarization fluctuations, breaks homogeneity in it and forms locally ordered regimes. Some of their effects are manifested in the dielectric behavior. As a result most of the 
disordered quantum paraelectrics show  a classical 
glassy behavior of a dipolar system, termed as relaxor\cite{relaxor}. In this work we focus on the effects of disorder in quantum critical behavior 
of certain ferroelectrics. Such issues were addressed in case of classical critical behavior earlier\cite{Dotsenko}. Also a quantum generalization of it in the context of itinerant magnets have been proposed in the recent 
past\cite{Millis}\cite{Jones}\cite{Vojta}. We use some of the earlier results and develop a new mean field description for possible low temperature dielectric behavior of a disordered quantum critical paraelectric. 
\section{Replica Mean Field Theory}  
Microscopically a quantum paraelectric is a system of interacting dipoles. If we consider dipoles as Heisenberg spins interacting with long range dipolar 
interactions, then the longitudinal modes are gapped out and  only transverse fluctuations are able to be soft\cite{Millis-1}. In experiments also we see only transverse optic modes to be soft during a system undergoing ferroelectric transition. Thus polarization fluctuations in these materials are well described by transverse optic mode fluctuations around zone center.
For simplicity, we restrict ourselves to one transverse  branch.  Assuming that the system is soft in one transverse  
direction, our effective action is an one component Landau-Ginzburg-Wilson action.  Near a ferroelectric quantum critical point it can be truncated at the lowest order and in a symmetric paraelectric  phase it takes the form  
\bea
\mA_{pure} &=& \frac{1}{2\beta}\sum_{n, \mb{q}}(\omega^2_n + \mb{q}^2 + r)|\phi(\omega_n, \mb{q})|^2+\frac{u}{4!}\int d\mb{x}\int_0^{\beta} d\tau\phi^4(x, \tau) .
\eea
Here $\phi(x,\tau)$ represents local polarization(fluctuating dipole) density at a position $x$ and at imaginary time $\tau$ in a coarse grained picture, while $ \phi(\omega_n, \mb{q})$ is it's Fourier transform. In usual notation, $\beta=1/k_BT$ where $k_B$ is the Boltzmann constant and $T$ being the temperature. Here $q$ is crystal momentum and $\omega_n=2\pi n T$ with  integer $n$, is the Bosonic Matsubara frequency. If we neglect the quartic term, i.e. in the non-interacting limit, propagator for polarization fluctuations in paraelectric phase $G^0(\omega, q)\sim \frac{1}{r+q^2-\omega^2}$. Here $-ImG^0(\omega,q=0)$ determines spectral function of the polarization fluctuations and in the $r=0$ limit it becomes gap-less. Thus the parameter $r$ determines the gap in polarization fluctuations. On the other hand starting from an ordered phase$(r<0)$, a mean field approximation
of the form $\phi(\omega, q)=\phi_0\delta(\omega)\delta(q)$ gives a Landau free energy density of the form
$f\sim r\phi^2_0+u\phi_0^4$. In this scheme $r=0$ is the point where $\phi_0$ vanishes. Thus it the is the mean field quantum critical point of the pure system. We consider vanishing spectral gap limit starting from a paraelectric phase i.e. $r\rightarrow 0^+$ and consider fluctuation corrections in presence of quenched disorder. Disorder is introduced into the problem as a random variation of $r$ in real space and is time independent. The disorder contribution to the above mentioned quantum action is given by,
\be
\mA_{dis} = -\frac{1}{2}\int d\mb{x}\int_0^{\beta} d\tau\delta r(\mb{x})\phi^2(x, \tau).
\ee 
Here $\delta r(\mb{x})$ is assumed to follow a Gaussian probability distribution  with variance $g$, viz.
\be
P(\delta r(\mb{x})) \propto \exp\{-\frac{1}{4g} \int d\mb{x}\delta r^2(\mb{x})\},
\ee

 so that $\overline{\delta r(\mb{x})} = 0$ and $\overline{\delta r(\mb{x})\delta r(\mb{y})} = g \delta^d(\mb{x}-\mb{y})$. 
To calculate the disorder averaged free energy, we introduce replicas of the order parameter $\phi_a$ with replica index a $= 1,...., n$ and  write $\overline{F}=-\frac{1}{\beta}(\overline{Z^n}-1)/n$, taking $n\longrightarrow0$ at the end of the 
calculation. Disorder averaging generates `non-local in time' interactions between fields with different replica indices and the resulting action becomes,
\bea
 \mathcal{A} &=& \frac{1}{2\beta}\sum_{m, \mb{q}, a, b}(\omega^2_m + \mb{q}^2 + r)|\phi_a(\omega_m, \mb{q})|^2 \del_{ab}+\frac{u}{4!}\sum_{ab}\int d\mb{x}\int_0^{\beta} d\tau\phi_a^4(x, \tau) \del_{ab}\nonumber\\
&&-\frac{g}{4}\sum_{ab}\int d\mb{x}\int_0^{\beta} d\tau\int_0^{\beta} d\tau^{'}\phi_a^2(x, \tau)\phi_b^2(x, \tau^{'}).
\label{A_rep}
\eea
Here $a, b$ are the replica indices which take positive integer values up-to some integer $n$. 

To begin with, we consider a replica symmetric case. We define replica symmetric solution as replica independent field configurations, i.e. 
\be
\phi_a(\mb{x}, \tau)= \phi(\mb{x}, \tau) \,{\rm for\, all}\,\, a
\ee
and their replica diagonal two point correlation functions, i.e.
\be
\chi_{ab}=\chi_{aa}\delta_{ab}=\chi_{0}\delta_{ab} \,{\rm for\, all}\, a,\, b.
\ee
 We consider a paraelectric phase i.e. $<\phi> =0$  and make a self-consistent quasi-harmonic approximation to decouple the quartic term. In this scheme a quartic term  is decomposed as 
 \be 
 \int dx d\tau \phi^4(x,\tau)\approx \lambda_0 \int dx d\tau \phi^2(x,\tau)),
 \ee  where $\lambda_0 =\int dx^{'}d\tau{'}\phi^2(x^{'},\tau{'})$. Thus susceptibility of the disordered paraelectric can be written as,
\be
\chi_0(\omega_m, q)= \frac{1}{(\omega^2_m + \mb{q}^2 + r+\lambda_0)}.
\ee
In the above equation, $\lambda_0$ describes the fluctuation corrections to ferroelectric gap and is defined by the following self-consistent equation
\bea
\lambda_0 &=& \sum_{m, \mb{q}} (u\chi_0(\omega_m, q)-g \chi_0(0, q))= u\int d^3 q \frac{1}{\Omega_q} \coth \beta\Omega_q-g \int d^3 q \frac{1}{\Omega^2_q}.
\label{self_sym}
\eea
The fluctuation corrected optical phonon frequency $\Omega_q$ is defined as,
\be \Omega^2_q = \mb{q}^2 + r+\lambda_0.
\ee 
It is to be noted that the second term in  eqn. (\ref{self_sym}) is a zero frequency contribution. The reason is that we consider quenched disorder which has no dynamics and thus strongly correlated in `time direction'. However the above two equations can be obtained by integrating $\delta r(x)$ without introducing replica trick and 
need to be solved self-consistently. It is clear form the expression for $\lambda_0$ that the second integral 
in the eqn. (\ref{self_sym}) gives a shift in gap and depending on its strength controls quantum fluctuations. In this scheme the solution of the equation $r-g \int d^3 q \Omega^{-2}_q= 0$ for $r$  gives the quantum critical point. Although a quantum critical point is strictly defined at zero temperature, it controls physical properties at finite temperature.\cite{book}  Right at that point various physical quantities at finite temperature follow  power law dependencies in temperature, e.g. static dielectric susceptibility $\sim T^{-2}$.
The effects of disorder considered here are identical to the effects of hydrostatic pressure as discussed in our previous work\cite{I}.

To include the effects of spatial inhomogeneity created by disorder, we need to break the replica symmetry at the vector level\cite{Dotsenko}.  In this approximation field configurations are assumed as
\bea
 \phi_a(\mb{x},\tau) &=& \phi_k(\mb{x}, \tau) +\psi(x,\tau)\, {\rm for}\, a=1, .., k \nonumber\\
 \phi_a(\mb{x}, \tau) &=&\psi(x,\tau)\, {\rm for}\, a =k+1, ....., n,
\eea 
and the correlation functions  to be block-diagonal 
\bea
\chi_{ab}(x,\tau) &=& \chi_{1}(x,\tau)+ \chi_{2}(x,\tau)\del_{ab}\, {\rm for}\, a,b=1,.., k\nonumber\\
&=& \chi_{2}(x,\tau)\del_{ab} \,{\rm for}\,a =k+1, ....., n.
\eea
%\begin{center}
 %\includegraphics[width=5cm]{matrix.eps}
%\end{center}
 Here $k\ge 1$ is an integer that determines the degree of the symmetry breaking process. 
In classical treatment for $gk>u$, $\phi_k$ corresponds to localized solution given as,
\be
 \phi(\mb{x}) = \sqrt{\frac{r}{\beta (gk-u)}} \psi(\sqrt{r}x).
\ee 
So that $\psi(z)$ obeys a scale independent equation
\be 
-\nabla_z^2 \psi(z) + \psi(z) - \psi^3(z) =0.
\label{scale-ind-eqn}
\ee 
The proper boundary conditions are $\psi(0) =$ constant and $\psi(\pm \infty)= 0$. 
Eqn. (\ref{scale-ind-eqn}) has exponential decaying solutions for $x\gg\sqrt{r} $ and is smooth for $x <\sqrt{r} $.
 The size of the droplet $ R$ is determined by the dipolar correlation length and $ R\sim \frac{1}{\sqrt{r}} $.
At very low temperature the dynamics of the droplets become important. In a simplest approximation spatial and the 
time dependent parts of the polarization field can be decoupled completely\cite{castro}. 
\be 
\phi(x, \tau) = \phi_k(x)\mT(\tau).
\label{RDA}
\ee 
Within this approximation the dynamics of the localized solution can be cast as a problem of an undamped Bosonic particle in a double well potential with frequency $\omega_0= 2$ in some dimensionless unit.  Authors of  reference\cite{castro} derived a tunnel splitting of ground state energy 
\be
r_L \approx 2e^{-r_0/r} 
\ee
with $r_0\sim E_1/E_2$  a constant, where $E_{N}=\int dz \phi(z)^{2N}$. Integrating out the fluctuations due to droplets in a Gaussian approximation, an effective action for the paraelectric fluctuations $\psi(x,\tau)$  can be written as, 
\bea
\mathcal{S}[\psi]= \frac{1}{2\beta}\sum_{m, \mb{q}, a, b}((\omega^2_m + \mb{q}^2)\del_{ab} + \mathcal{M}_{ab})\psi_a\psi_b.
%\nonumber\\
%&+&\frac{u}{4!}\int d\mb{x}\int_0^{\beta} d\tau\psi_a^4(x, \tau) \del_{ab}\nonumber\\
%&-&\frac{g}{4}\int d\mb{x}\int_0^{\beta} d\tau d\tau^{'}\psi_a^2(x, \tau)\psi_b^2(x, \tau^{'}).
\eea
The presence of droplets introduces a ``gap-matrix'' $\{\mathcal{M}_{ab}\}$ which contains $k\times k$ 
block with elements,
\bea
\mathcal{M}_{ab} = r(1-\frac{g k-3u}{gk-u}\lambda_L)\delta_{ab}-\frac{2g kr}{gk-u}\lambda_L
\label{gap_k}
\eea
and diagonal elements for the remaining $n-k$ replicas
\bea
\mathcal{M}_{ab} = r(1-\frac{g k}{gk-u}\lambda_L)\del_{ab}.
\label{gap_n}
\eea
%In case of $r_L\rightarrow 0$ limit, because of large tunneling the ferroelectric order in the locally ordered regime gets destroyed. However contributions from the fluctuating localized solutions remain. Fluctuation contribution to the effective action from the tunneling dynamics is encoded in 
Here $\lambda_L$ encodes the contributions from the localized solutions along with their dynamics and is given as
\bea
\lambda_L &=& \sum_{\omega}\int dz <\psi(z)\mathcal{T}(\omega)\psi(z)\mathcal{T}(\omega)>\nonumber\\
&=& \int dz\psi^2(z) \sum_{\omega}<\mathcal{T}(\omega)\mathcal{T}(\omega)>\nonumber\\
&\sim& \frac{1}{\omega_{-}} \,{\rm at}\, \, T=0.
\label{Lambda_l}
\eea
Here $\omega_{\pm} =2\pm r_L$. It is to be noted that the vector breaking of replica symmetry not only
introduces inhomogeneous solutions but also glassy effects through off-diagonal elements in the gap-matrix.
Putting $\lambda_L=0$ identically, we get back the behavior of a pure system.
However in this scheme replica correlators for disordered paraelectric is given by
\bea
\chi_{ab}^{-1}(\omega_m, q)= ((\omega^2_m + \mb{q}^2)\del_{ab} + \mathcal{M}_{ab}).
\eea
%Fluctuation correction along with self-consistency condition is encoded in $\lambda_{ab}$ and is given by
%\bea  \lambda_{ab} = u\sum_{m, q} \chi_{ab}(\omega_m, q)-\frac{g}{4}\sum_{q} \chi_{ab}(0, q)
%\eea
We look for dielectric instability points, i.e. where $\chi_{ab}^{-1}(0,0)=0$ using  equations (\ref{gap_k}) and (\ref{gap_n}). Diagonalized form of the gap-matrix is given as,
\bea
\hat{\mathcal{M}}_{aa} = 
\begin{cases}r(1-\frac{g k-3u}{gk-u}\lambda_L),\, &a=1,..k-1,\\
r(1-\frac{3g k-3u}{gk-u}\lambda_L), \, &a=k,\\
r(1-\frac{g k}{gk-u}\lambda_L), &a=k+1,..,n.
\end{cases}
\label{Taa}
\eea
Using equations (\ref{Taa}) and (\ref{Lambda_l}) we find the values of $r$ at which zero temperature diagonal susceptibility $(\sim \frac{1}{\hat{\mathcal{M}}_{aa}})$ diverges. The  instability point depends on the disorder strength and the value of $k$ and is given as,
\bea
r_c=\begin{cases}
 -r_0/\log(1-A\frac{(g k-3u)}{(g k -u)}), \,\, &a=1,..,k\\
-r_0/\log(1-A\frac{(3g k-3u)}{(g k -u)}), \,\, &a=k\\
-r_0/\log(1-A\frac{g k}{(g k -u)}), \,\, &a=k+1,..,n
\end{cases}.
\label{rc}
\eea and is $k$ dependent. Here $A$ is a system dependent parameter. We can use this equation to describe a schematic phase diagram as shown
in Fig. 1.
\figph
 For a simple minded analysis, let us consider the $a=k+1,..,n$ elements only. For $|\frac{Ag k}{(g k -u)}|<<1$, $r_c$  can be written in the following form
\bea
 r_{c}(k)&=&r_0^{'}\frac{(g k -u)}{g k}\nonumber\\
&=& B-\frac{C}{k}\,\, (B=r_0^{'}, \, C=\frac{r_0^{'}u}{g}).
\eea 
 Since the choice of $k$ is random, depending on its distribution at the limit $n\rightarrow 0$, we can estimate a distribution, hence width of $r_c$. In a replicated action with $n$ replicas, $k$ can be chosen in $C^n_k$ ways. Thus we can define a normalized `distribution' of $\mathcal{P}(k)$ as follows
\be
 \mathcal{P}(k)=\frac{C^n_k}{\sum^n_{k=1} C^n_k }=\frac{1}{2^n-1} \frac{\Gamma(n)}{\Gamma(k)\Gamma(n-k)}.
\ee 
Since the gamma function with negative argument is infinity, the limit of $k$ can be extended to infinity. In the limit $n\rightarrow 0$, using the asymptotic form of gamma functions, $\mathcal{P}(k)$ can be approximated as\cite{Dotsenko}
\be
 \mathcal{P}(k)\approx \frac{1}{\log{2}}\frac{(-1)^{k-1}}{k}\approx-\frac{1}{\log{2}}\frac{\cos\pi k}{k}.
\ee
It is to be noted that $\mathcal{P}(k)$ is not a distribution function of any physical variable. It has negative values of $\mathcal{P}(k)$ for some values of $k$.  There are several possible broken replica symmetric cases, each characterized by the number $k$ which follows a distribution $\mathcal{P}(k)$. For a fixed disorder strength $g$, each $k$ results a different instability point $r_{c}$. Instead of $k$, if we characterize various possible broken replica symmetric cases by $r_{c}$, a distribution of $r_{c}$ can be estimated as
\be
 \mathcal{P}(r_{c}) =  \mathcal{P}(k) |\frac{\Delta k}{\Delta r_{c}}|\sim \frac{1}{B-r_{c}}\times \cos(\pi k).
\label{Pr2}
\ee
This is a broad power-law distribution of $r_{c}$ around a system dependent parameter $B$ with a cosine factor. %Inhomogeneous solutions in the vector breaking of replica scheme are stable for $k>u/g$. Thus for large
%disorder strength compared to anharmonicity, i.e. for small $u/g$ the droplet contributions are dominated by small $k$'s. 
The probability distribution can be assumed to be smooth around $k=$ any positive integer, excluding zero. 
The expansion around $k=0$ is excluded as it corresponds to small $u/g$ limit where the action (eqn. \ref{A_rep}) becomes unstable even in a replica symmetric ansatz. In that limit the system will undergo a first order transition in a replica symmetric analysis, the stability of the system needs a $\phi^6$ term in the action (eqn. \ref{A_rep}) which will lead to more complicated localized solutions in a broken replica symmetry picture. However we focus on those $u/g$ values where the above possibilities are not present and the distribution function is smooth.
%In the opposite limit, i.e. for large $u/g$ it becomes rapidly oscillating. We restrict to the small $u/g$ case for the rest of the discussion.  
It is to be noted that the power law nature of $\mathcal{P}(r_{c})$ arises because of the dynamics of the locally ordered regimes and also depends on the distribution of $k$ used.
Neglecting cosine factor within some range of $r_c$ say $(B+R, B-R)$, average susceptibility of the disordered quantum paraelectric can be estimated as,
\bea
\overline{\chi(r, T)}&\sim& \int_{B-R}^{B+R} dr_c \frac{1}{B-r_c}\times \frac{1}{r-r_c+T^2}\nonumber\\
&=& \frac{1}{r-B+T^2} \log\frac{r-B-R+T^2}{r-B+R+T^2}.
\label{chi_av}
\eea
It is evident that
inclusion of fluctuations due to locally ordered regime introduce a parameter $R\sim \mathcal{O}(u/g)$ and changes the usual quantum critical behavior of a paraelectric. In the limit $r\rightarrow B$, the temperature dependence of a disordered quantum paraelectric can be predicted as,
\bea
\overline{\chi(r, T)}\sim
\begin{cases}
{\rm constant,}\,\, T<< R\\
1/T^4, \,\, T>> R.
\end{cases}
\label{chi_av_asym}
\eea
 This is a deviation from the standard quantum critical behavior which predicts $\chi(T)\sim T^{-2}$ over whole temperature range in a pure system. Above estimate is somewhat limited as we are unable to estimate $\mathcal{P}(k)$ for all values of $k$ here. Replacement of summation over $k$ by an integral within a range defined
by the parameter $R$ introduces an artificial logarithmic divergence at $T=\sqrt{R}$. Within that, our analysis produces some features of relaxor behavior as shown in experiments on doped quantum paraelectrics. Now we follow references \cite{expt1}\cite{expt2} to make connection between our result with experiments. In these references,
we see variety of dielectric constant vs temperature curves, depending on doping concentrations. These results are
non-universal and they must depend on sample preparation. This makes a theory difficult to have perfect match with experiments in any disordered system. Here we consider the gross features of these experiments. We see, starting from a constant value at low temperature, static dielectric constant of disordered SrTiO$_3$ peaks at an intermediate temperature and falls thereafter. This is what is termed as `relaxor' behavior. The constant low temperature part and the high temperature tail for dielectric constant are produced here. The rounded peak behavior can be obtained if we overcome the limitations as mentioned above and left as a future task.  To summarize, our analysis predicts a relaxor type behavior with non-Curie-Weiss tail for static dielectric constant at high temperature.

If we consider other thermodynamic quantities, e.g. specific heat, then we see that in this model it will still be gapped. 
It follows from the following argument. In a pure gapped system, specific heat at low temperature has exponential dependence on spectral gap.
For a disordered system such a spectral gap has a distribution and an averaging over such a distribution can lead to
different temperature dependence. To get rid of the exponential dependence and to achieve a power law quantum Griffiths behavior\cite{book}, we need an exponential distribution of spectral gap which is absent in our study. Thus our model study predicts a gapped behavior in the low temperature regime.  It is neither a pure quantum critical nor a quantum Griffith like behavior. It is a general mixed phase behavior. 
\section{Discussions}
Low temperature dielectric behavior of a  quantum paraelectric in presence of quenched disorder is addressed here. A suitable action for these materials with random T$_c$ type disorder have been studied using a replica trick. The effects of disorder induced locally ordered regimes and their tunneling at low temperature are captured in this formalism. We derive an expression for the distribution of instability points for a fixed value of disorder strength and demonstrate the possibility of a mixed phase at any non-zero disorder strength. This analysis shows that the instability points follows a broad power law distribution around a system dependent parameter with a cosine correction. Using such distribution we are able to show analytically how the temperature dependence of static dielectric susceptibility of a disordered quantum critical paraelectric deviates from its pure counterpart. Our analysis is a completely new attempt in the context of the effects of disorder in ferroelectrics near 
a quantum critical point. In a qualitative manner it  predicts certain new features such as occurrence of a phase with mixture of critical and non-critical regimes with a distributions of transition points. This is  in contradiction with earlier numerical works in similar issues in context of  disordered spin systems\cite{1}\cite{2} which predicts conventional critical behavior. Considerations of different glassy correlations in different studies result
in such differences. We consider it through replica symmetry breaking.  On the other hand authors in the references 
\cite{1}\cite{2} assume disorder effects in 2D spin systems maintaining  dimer correlations introduced a priori followed by a numerical analysis. Exact reason behind such differences are not clear to us and is most likely those studies resemble different systems. Also, our study is a mean field description which neglects some of the
fluctuation effects.   In spite of few limitations the whole analysis is interesting in context of the use of replica trick to incorporate disorder induced inhomogeneities or locally ordered regime along with their dynamics in the studies of quantum phase transition and may turn out to be useful in explaining certain experimental results on disordered ferroelectrics near a quantum critical point.
\section*{Acknowledgment} 
The author would like to thank S M Bhattacharjee and S G Mishra for many useful discussions.

\end{document}